\documentclass[aps,prl,twocolumn,groupedaddress]{revtex4}

\usepackage{graphicx}
\usepackage{dcolumn}

\begin{document}


\title{Effective field theory approach to electroweak transitions of nuclei far from stability}


\author{M. A. Huertas}
\affiliation{College of William and Mary, Williamsburg, VA}


\date{\today}

\begin{abstract}
In a previous paper, the convergence of the effective field theory approach of Furnstahl, Serot and Tang to the nuclear many-body problem was studied by applying it to selected doubly-magic, and neighboring single-particle and single-hole, nuclei far from stability. The success of that approach, interpreted through density functional theory, would imply reliable densities. In this paper, the single-particle (Kohn-Sham) wave functions are probed using weak transitions near the Fermi surface. The weak currents are the Noether currents derived from the effective Lagrangian. The general single-particle transition matrix elements, from which any semi-leptonic weak rate can be calculated, are obtained in terms of upper and lower components of the Dirac wave functions. Here beta-decays in nuclei neighboring $^{132}$Sn are studied and compared with available experimental data. Calibration of the theoretical results for such decays may also have useful application in element formation.
\end{abstract}


\maketitle


\section{Introduction}

In a recent paper~\cite{huertas} the convergence of calculations to experimental results at different levels of approximation from a new approach to the nuclear many-body problem was studied. This approach combines the principles of Effective Field Theory (EFT) with Density Functional Theory (DFT). The results of that work showed that the total binding energy of even-even Sn isotope nuclei can be reproduced below the 1\% level. In addition to this agreement for the total binding energy of the doubly-magic nuclei $^{132}_{\ 50}$Sn$_{82}$, $^{100}_{\ 50}$Sn$_{50}$, $^{78}_{28}$Ni$_{50}$ and $^{48}_{28}$Ni$_{20}$, the chemical potential for neighboring nuclei, differing by one particle or hole from the doubly-magic ones, was also well reproduced below the 10\% level.
The agreement in binding energies shows that the energy functional derived from the effective lagrangian of Furnstahl, Serot and Tang \cite{furnstahl} is indeed a good approximation and thus, according to DFT, the ground-state densities obtained in each case are also a good approximation to the true ground-state densities. Although in~\cite{huertas} both proton and neutron densities for $^{132}$Sn and $^{100}$Sn were presented, there has been no direct measurement of either of them and thus the comparison with experiment has not yet been established directly. 

In the Kohn-Sham approach, the ground-state density is constructed from single-particle wave functions, obtained by solving the Kohn-Sham equations of the system. These equations are the energy eigenvalue equations for a system of non-interacting particles subject to a local external potential. Except for the energy eigenvalue close to the Fermi surface (i.e. the energy necessary to extract one particle from the system) all other energy eigenvalues have no direct physical meaning. If the single-particle (hole) energy at the Fermi surface agrees with experiment, one can assume that the wave function associated with it must also be a good approximation.
In this paper the validity of the last statement is studied, and in an indirect way the accuracy to which the ground state density is reproduced. The wave functions describing one particle outside a doubly-magic core or one hole in a doubly-magic core are used to calculate various $\beta$-transition rates. These wave functions represent the initial or final nuclear states (in coordinate representation) in these transitions. Here, based on the above arguments, an accurate description of the wave function is assumed to be guaranteed by the accuracy with which its binding energy is reproduced. Therefore the ideal case of study corresponds to a ground-state to ground-state transition, since in these cases there is a closer agreement with the experimental energy values. Wave functions of excited states will be less accurate as these states lie farther away from the Fermi surface and thus their energy is not well reproduced. Furthermore, it is expected that the particle-particle transition will give a cleaner result since the description of a particle outside the doubly-magic core can be well approximated by a single-particle wave function. The description of a hole, on the other hand, is more complicated since it represents a more complex many-body state and its description by a single-particle wave function might be expected to be less accurate. 
Using the results obtained in~\cite{huertas} and the available experimental data, this paper focuses on the core nucleus $^{132}_{\ 50}$Sn$_{82}$ and its single-particle and single-hole neighbors $^{133}_{\ 50}$Sn$_{83}$, $^{131}_{\ 50}$Sn$_{81}$, $^{133}_{\ 51}$Sb$_{82}$ and $^{131}_{\ 49}$In$_{82}$. Using these nuclei, the two types of transitions, particle-particle and hole-hole, can be investigated and compared with experimental results to validate the observations made above. The first case, i.e. the particle-particle transition, corresponds to the $\beta$-decay process $^{133}_{\ 50}$Sn$_{83}$ $\rightarrow$ $^{133}_{\ 51}$Sb$_{82}+e^{-}+\overline{\nu}_{e}$ and the second one, the hole-hole case, to the transition $^{131}_{\ 49}$In$_{82}$$\rightarrow$$^{131}_{\ 50}$Sn$_{81}+e^{-}+\overline{\nu}_{e}$. In both cases ground-state to ground-state transitions, as well as transitions from and to low-lying excited states, are considered. 

To study these processes a general expression describing semi-leptonic transitions has been obtained that incorporates the Dirac wave functions calculated by solving the Kohn-Sham equations derived from the effective lagrangian~\cite{huertas,furnstahl}. The upper and lower components of the Dirac spinors are then used to calculate the matrix elements of the appropriate electroweak currents. These currents are obtained from the same effective lagrangian and correspond to the leading order Noether currents~\cite{currents,waleckabook}. At any level of approximation in the effective lagrangian, the axial-vector currents satisfy PCAC and show a pion-pole dominance~\cite{currents}. The matrix elements of the currents thus constructed are used in a general multipole expansion from which \textit{any semi-leptonic weak interaction} can be calculated~\cite{waleckabook}. This paper focuses on $\beta$-decay rates. Applications to other semi-leptonic processes are being considered for future work. The calculated decay rates have been corrected for the screening effect from the valence electrons in the daughter nucleus~\cite{landolt} and for the slowing down of the emitted electrons due to the attractive Coulomb core (Fermi function). This last correction uses a relativistic description of the electron and takes into account the effects of the size of the nucleus~\cite{landolt}.

Calculations of $\beta$-decay properties, like the half-live ($\tau_{1/2}$) and $\beta$-delayed neutron emission (P$_{n}$), have been done extensively using mainly the following models: gross theory, quasi-particle random phase approximation (pn-QRPA) and shell model calculations. Gross theory has been used in large-scale calculations where the discreetness of the final energy levels of the daughter nucleus are smoothed out and different single-particle strength functions (Gaussian, modified Lorentz) are used to calculate the $\beta$-decay strength~\cite{gross}. This approach has been used to calculate both allowed and first-forbidden transitions using the Q-values from mass formulas as input. Improvements to this model have been made in which pairing and other shell effects are taken into account~\cite{newgross}. Shell model calculations are one of the most elaborate methods used. In this type of calculation it is possible to incorporate multi-particle transition amplitudes, which are uniquely determined by the specification of the model Hamiltonian. Such an approach has the advantage of predicting the state and mass dependence of observed decay strengths making the results independent of any mass formulas input. The disadvantage of this method lies in the large matrices that have to be computed when the number of nucleons increases, therefore its applications has been limited to small nuclei.  A calculation for proton-rich nuclei has been attempted for the cases of $sd$-shell nuclei and can be found in~\cite{shell}. The third approach, pn-QRPA, has also been used in extensive calculations of $\beta$-decay observables. This approach can be considered to be between the shell model and gross theory. A description of the formalism for even-even mother nuclei, as well as references, can be found in~\cite{qrpa}. The extensions to odd systems and odd nuclei can be found in~\cite{qrpa-o} and~\cite{qrpa-oo}.

From the above models used to calculate $\beta$-decay half-lives, QRPA has shown to give good agreement with experimental results. Some extensive calculations have been performed~\cite{staudt-data-tables,bender,staudt-z} showing an agreement with experiment within a factor of two.
The calculations included in~\cite{staudt-data-tables} 
describe the $\beta$-strength function by applying the pn-QRPA method with a Gamow-Teller residual interaction, the strength of which is fitted to experimentally known half-lives of known isotopes for a fixed mass number A. Pairing correlations are treated in the BCS model using a constant pairing force and without taking into account the Pauli blocking. The proton and neutron gaps are equal and are taken from the values of global systematics. In addition to this, spin-isospin ground-state correlations are included. For very neutron-rich nuclei only allowed Gamow-Teller transitions are considered and the influence of first-forbidden transitions on the half-lives of nuclei far from stability is neglected. Nuclear deformations also are taken into account using the Nilsson model from which the wave functions of the parent and daughter nuclei are calculated assuming the same ground-state deformation for both. The main uncertainties in these calculations come from the mass formulas used as input, which in general become less accurate for nuclei far from stability. The results obtained show an agreement with experimental data for nuclei with short half-lives ($<$1 s) with an average deviation of 1.4. 

Other calculations have concentrated on some specific range of nuclei important for the r-process such as those close to $^{132}$Sn, as in our case. In~\cite{borzov} 
an analysis of the ground-state properties and $\tau_{1/2}$ of nuclei close to $^{132}$Sn is performed using the Hartree-Fock-Bogoliubov plus BCS pairing approach (HFB+BCS). Although some studies indicate that the nuclei in this region are spherical, the calculations included both spherical and deformed nuclei. Contributions from pairing using a constant strength, zero-range force were also included. Again here, only allowed GT transitions were calculated and the GT strength function was obtained using the method of a self-consistent treatment of the ground and excited states of even-even and odd-A superfluid nuclei solving QRPA-like equations in the finite Fermi-system (FFS) theory. Using this method, the Q$_{\beta}$ were also determined and no additional mass formulas were used. A density functional describing the nuclear system was used with parameters fitted to stable nuclear properties obtaining three different sets. One of them was specialized to reproduce not only known ground-state properties of magic nuclei but also single particle energies of $^{132}$Sn.  Details and additional references are given in~\cite{borzov}.
A similar study~\cite{moeller},
focused on nuclei in the r-process path, used a general method that combines the microscopic QRPA model for allowed GT $\beta$-decay with statistic gross theory of first-forbidden decays. In general, the results show that there is a much better agreement with experiment when the first-forbidden transitions are included, especially for large values of $\tau_{1/2}$. When the calculated half-lives for a large range of nuclei were compared with experimental results it was found that the quantity $ln(\tau_{calc}/\tau_{exp})$ lies in the range 10-0.1.

Because of new data and improved methods to calculate $\beta$-decay properties, there have been efforts to compile the existing experimental, as well as calculated data. In~\cite{pfeiffer} 
such a compilation can be found which includes the most recent experimental data for half-lives as well as the results from two different models: 1) Kratz-Herrmann formula (KHF) and 2) Macroscopic-microscopic QRPA. 

\begin{table*}
\caption{\label{tb: calc-half-lives} Calculated $\beta$-decay half-lives (in ms) of nuclei close to $^{132}$Sn.}
\begin{ruledtabular}
\begin{tabular}{r|rrr|rrrrr|rr|r}
Nuclei & \footnotemark[1]KHF &\footnotemark[1]QRPA-1& \footnotemark[1]QRPA-2 & \footnotemark[2]DF2& \footnotemark[2]DF3 & \footnotemark[2]R1 & \footnotemark[2]R2 & \footnotemark[2]R3 &\footnotemark[3]GT &\footnotemark[3]GT+ff &\footnotemark[4]exp\\
\hline
$^{133}_{\ 50}$Sn$_{83}$&362&9479&9479&9320&8200&823&10290&1260&- &-&1450 $\pm$ 30\\
$^{131}_{\ 49}$In$_{82}$&216&146&146&390&350&394&1470&332&147.1 &139.2&280 $\pm$ 30\\
\end{tabular}
\footnotetext[1]{See~\cite{pfeiffer}.}
\footnotetext[2]{See~\cite{borzov}. }
\footnotetext[3]{See~\cite{moeller}.}
\footnotetext[4]{See~\cite{expdata-sn133,expdata-in131}.}
\end{ruledtabular}
\end{table*}

%

Table~\ref{tb: calc-half-lives} summarizes the results obtained from the methods discussed above in the particular cases of the nuclei $^{133}$Sn and $^{131}$In. Here one can appreciate the level of agreement with experiment. The best agreement for the decay of $^{133}$Sn is given by R3, with a ratio of calculated to experimental half-lives, $\tau_{1/2}$, equal to 0.87 or a deviation of 13\%. In the case of $^{131}$In, the best agreements are obtained by KHF and R3 with a deviation of 22\% and 18\%, respectively.

All these models aim to reproduce the known experimental $\beta$-decay half-lives as well as predict its value for other nuclei, especially those far from stability. As pointed out in~\cite{pfeiffer}, most of the models can be grouped in two categories: those which give a mathematical expression (e.g.~a polynomial) for the quantity of interest and those based on an effective interaction. Models of the first type have no direct link to the underlying nucleon-nucleon interaction and do not give additional information regarding the nuclear single-particle wave functions. For models of the second type, these approaches use different effective interactions fitted to better reproduce the experimental $\beta$-decay data. In other cases, because there is an overparametization of the effective interaction, its relation with the nucleon-nucleon interaction is also lost. Most of these models combine different approaches and approximations to better reproduce the experimental data.

In general, it can be said that there is \textit{no consistent theory that by fixing its parameters once, reproduces not only the masses but also the single-particle levels and $\beta$-decay properties of nuclei}.


In this paper such a consistent approach is explored and compared with experimental data for $\beta$-transition rates of neighboring nuclei $^{132}$Sn . This approach corresponds to a new theory that combines Density Functional Theory and Effective Field theory, where an energy functional is constructed which is consistent with the symmetries of QCD and whose parameters are fitted to properties of stable nuclei~\cite{furnstahl}. The single-particle wave functions are obtained by solving the self-consistent Kohn-Sham equations. In order to maintain a consistent approach, the correct electroweak currents, needed to calculate the $\beta$-transition rates, are derived from the same effective lagrangian. These currents are the leading order Noether currents. The axial-vector current satisfies PCAC at any order in the effective lagrangian~\cite{currents}. A comparison with experimental data will give this approach its validation and limitations as a calculational tool and will also test implications of DFT regarding the single-particle wave functions used to construct the ground state density in the Kohn-Sham approach. 

The results for the ground-state particle-particle $\beta$ transition rate show an excellent agreement with experiment (to 5\% in the case of the ground-state to ground-state transition from $^{133}$Sn to $^{133}$Sb). This is in agreement with the expectation that since the binding energy of the nuclei and the single-particle energy close to the Fermi surface of the particle outside the core are well reproduced, then the single-particle wave-function are also well reproduced. Transitions to excited states show a systematic deviation from experimental data, again in agreement with the expectation about the wave functions of those excited states. In the case of hole-hole transitions, studied here, the results are less reliable.

The paper is organized as follows: in section 2 a general expression for semi-leptonic transition rates in terms of a multipole expansion and using Dirac wave functions is presented; in section 3 this formulation is used to calculate $\beta$-transition rates for selected nuclei and results are compared with the existing experimental data; and finally, section 4 contains the conclusions drawn from this study.

\section{General Semi-leptonic processes}

The calculation of $\beta$-decay rates in this paper is based on a general expression that can be applied to any semi-leptonic process. In this section the derivation of that expression is discussed. Much of the material presented here follows~\cite{waleckabook}.

The starting point is the interaction hamiltonian $H_{\rm w}$ which for low energy processes, like $\beta$-decay, is described by the semi-leptonic weak hamiltonian of the standard model. This interaction hamiltonian is written down to first order in the weak constant, G. This implies that leptons are treated to this order but the strong interaction part of the hamiltonian is treated to all orders. The interaction hamiltonian is described by a current-current form \footnote{For charge-changing semileptonic processes G$^{\pm}$=G$cos \theta_{C}$ where G=1.0267x10$^{-5}$/$m^{2}_{p}$ and $cos \theta_{C}=0.974$. See~\cite{waleckabook}}
\begin{equation}
\label{inthamiltonian}
H_{\mathrm{w}}= -\frac{G}{\sqrt{2}} \int d^{3}x j_{\mu}^{\mathrm{lept}}(\vec{x})\mathcal{J}_{\mu}(\vec{x})
\end{equation} 
where $j_{\mu}^{\mathrm{lept}}(\vec{x})$ is the lepton current and $\mathcal{J}_{\mu}(\vec{x})$ the hadronic current. By taking matrix elements of this hamiltonian between the initial and final states (lepton and hadron) the leptonic and hadronic parts factorize and the matrix elements of the leptonic currents can be expressed by
\begin{equation}
\langle f_{l}|j_{\mu}^{\mathrm{lept}}(\vec{x})|i_{l} \rangle = l_{\mu}e^{-i\vec{q}\cdot \vec{x}}
\end{equation}
where $\vec{q}=\vec{k}_{\rm{e}}+\vec{k}_{\nu}$ is the momentum transferred in the process and $\vec{k}_{\rm{e}}$, $\vec{k}_{\nu}$ are the corresponding electron and neutrino momenta. $f_{l}$ and $i_{l}$ represent initial and final lepton states. In addition, $l_{\mu}=(\vec{l},il_{0})$ and $l_{3}=\vec{l} \cdot \hat{q}$, where $\hat{q}$ is a unit vector in the direction of $\vec{q}$, i.e. $\vec{q}/|\vec{q}|$. 

Using the expression in Eq.(\ref{inthamiltonian}) and making a multipole expansion of the hadronic current to project irreducible tensor operators, one arrives at the following expression for the matrix elements~\footnote{Single nucleon matrix elements of the currents also contain form factors~\cite{furnstahl,waleckabook} which are unimportant for the low momentum transfer processes considered here.}. In this and the following equations the magnitude of $\vec{q}$ is defined by $k=|\vec{q}|$.
\begin{widetext}
\begin{eqnarray}
\label{me-int-h}
\langle f|H_{\mathrm{w}}|i \rangle = \frac{-G}{\sqrt{2}} \langle f| \{ -\sum_{J\geq 1} \sqrt{2\pi (2J+1)} (-i)^{J} \sum_{\lambda=\pm 1} l_{\lambda} \left[ \lambda \mathcal{T}_{J-\lambda}^{\mathrm{mag}}(k)+\mathcal{T}_{J-\lambda}^{\mathrm{el}}(k) \right]
\nonumber \\
 + \sum_{J\geq 0} \sqrt{4\pi (2J+1)} (-i)^{J} \left[l_{3} \mathcal{L}_{J0}(k)-l_{0}\mathcal{M}_{J0}(k) \right] \}|i \rangle
\end{eqnarray} 
\end{widetext}
The four multipole operators occurring in the previous expression are defined in the following way~\cite{waleckabook,donnelly-1}:
\begin{equation}
\label{eq:multipole-M}
\mathcal{M}_{JM}(k) = \int d^{3}x \left[ j_{J}(kx)Y_{JM}(\Omega_{x}) \right] \mathcal{J}_{0}(\vec{x})
\end{equation}
\begin{equation}
\label{eq:multipole-L}
\mathcal{L}_{JM}(k) = \frac{i}{k} \int d^{3}x \{ \vec{\nabla}\left[j_{J}(kx)Y_{JM}(\Omega_{x}) \right] \} \cdot \vec{\mathcal{J}}(\vec{x})
\end{equation}
\begin{equation}
\label{eq:multipole-T-ele}
\mathcal{T}^{\mathrm{el}}_{JM}(k) = \frac{1}{k} \int d^{3}x \{ \vec{\nabla} \times \left[j_{J}(kx)\vec{Y}_{JJ1}^{M}(\Omega_{x}) \right ] \} \cdot \vec{\mathcal{J}}(\vec{x})
\end{equation}
\begin{equation}
\label{eq:multipole-T-mag}
\mathcal{T}^{\mathrm{mag}}_{JM}(k) = \int d^{3}x \left[j_{J}(kx)\vec{Y}_{JJ1}^{M}(\Omega_{x}) \right] \cdot \vec{\mathcal{J}}(\vec{x})
\end{equation}

The multipole operators have the vector--axial-vector (V-A) structure. The general form of the hadronic current is given by
\begin{equation}
\mathcal{J}_{\mu} = \mathrm{J}_{\mu}+\mathrm{J}_{\mu 5}
\end{equation}
and the different parts of the total current $\mathcal{J}_{\mu}$ (vector and axial) used in this paper are given by

\begin{equation}
\label{currents}
J_{\mu} = i\psi^{\dag}\gamma_{4}\gamma_{\mu}\psi + \frac{(\lambda_{p}-\lambda_{n})}{2m} \frac{\partial}{\partial x_{\nu}}(\psi^{\dag}\gamma_{4}\sigma_{\mu \nu}\psi)
\end{equation}
\begin{equation}
\label{currents-axial}
J_{\mu 5}= \left(\delta_{\mu \nu} + \frac{1}{m^{2}_{\pi}-\partial_{\alpha} \partial_{\alpha}} \frac{\partial}{\partial x_{\mu}}\frac{\partial}{\partial x_{\nu}}
\right) F_{A}i\psi^{\dag}\gamma_{4}\gamma_{5}\gamma_{\nu}\psi 
\end{equation}
where $\lambda_{p}$ and $\lambda_{n}$ are the anomalous magnetic moments of the nucleon and $\partial_{\alpha} \partial_{\alpha}$ stands for the D'alambertian operator. $F_{A}$ is taken as a constant, i.e. we consider here small momentum transfers, and its numerical value is -1.257. These currents are the leading currents as described in~\cite{furnstahl,currents}. There are no additional contributions to the one-body current coming from higher order terms in the effective lagrangian~\cite{currents}. The next contribution to the current comes from two-body effects which are estimated to be of the order of $O(q/m_{\pi})$, with $q$ the momentum transfer. Since the maximum value of $q$ is of the order of a few MeV, this contribution corresponds to a few percent correction. Additional contributions, from still higher order terms, are reduced by a factor of $q/M$, with $M$ the nucleon mass. These estimates do not take into account the spin structure of these two-body currents, and the actual correction will depend on a detailed calculation, which lies beyond the scope of the present work~\footnote{See for example~\cite{ananyan}}.

In order to evaluate the nuclear matrix elements of the four multipole operators it is necessary to define the basis wave functions that are going to be used. For this we employ the wave functions obtained by solving the Kohn-Sham equations corresponding to the effective lagrangian that describes the nuclear many-body system. This lagrangian and the corresponding Kohn-Sham equations can be found in ~\cite{furnstahl,huertas}.

At this point, the important aspect of the wave functions used is that they are described by Dirac spinors and thus the contributions of both upper and lower components have to be taken into account to all orders. Both currents and wave functions are used without a non-relativistic reduction. The form of the Dirac spinors is given by
\begin{equation}
\label{dirac-spinors}
\psi_{n \kappa m}(\vec{x}) = \frac{1}{r} \left( 
\begin{array}{r}
iG(r)_{n \kappa}\Phi_{\kappa m}(\Omega_{x}) \\
-F(r)_{n \kappa} \Phi_{-\kappa m}(\Omega_{x}) \\
\end{array}
\right) \eta_{t}
\end{equation} 
where $\Phi_{\kappa m}(\Omega_{x})$ are spin spherical harmonics defined by 
\begin{equation}
\Phi_{\kappa m}(\Omega_{x}) = \sum_{m_{l}m_{s}} \langle lm_{l}\frac{1}{2}m_{s}|l\frac{1}{2}jm \rangle Y_{lm_{l}}(\theta , \phi) \chi_{m_{s}}
\end{equation}
and $\chi_{m_{s}}$ are two-component spinors. $\eta_{t}$ is a two component isospinor describing either a proton or a neutron~\footnote{The quantum number $\kappa$ is related to the orbital angular momentum quantum number, $l$ by the following relation: $l=\kappa$ if $\kappa > 0$, $l=-(\kappa + 1)$ if $\kappa < 0$. The magnitude of $\kappa$ is given by $|\kappa|=j+1/2$, where $j$ is the total angular momentum quantum number ~\cite{waleckabook}.}.

The currents as well as the multipole operators have a matrix structure (two by two block matrix) due to the Dirac matrices included in them, mixing the upper and lower components of the initial and final nucleon wave functions. In what follows, the elements of the matrix form of these multipoles play an important role and to keep track of them they will be labeled as follows.
\begin{equation}
\label{multipole-matrix}
O_{JM}=\left (
\begin{array}{cc}
d_{1}^{JM} & b^{JM} \\
b^{JM} & d_{2}^{JM} \\
\end{array}
\right )
\end{equation}
Using Eqs.(\ref{currents}) and (\ref{currents-axial}) and the following matrix form for the Dirac $\gamma$ matrices (here the conventions of \cite{waleckabook} are used)
\begin{equation}
\vec{\gamma} = \left (
\begin{array}{cc}
0 & -i\vec{\sigma} \\
i\vec{\sigma} & 0 \\
\end{array}
\right )
\end{equation} 

\begin{equation}
\beta = \left (
\begin{array}{cc}
1 & 0 \\
0 & -1 \\
\end{array}
\right )
\end{equation} 
then the elements of the multipole matrices can be obtained.

The matrix elements of the multipole operators shown in Eq.(\ref{me-int-h}) will mix the upper and lower components of the Dirac wave functions as shown in the following expression. Here all quantum numbers referring to the final state are indicated by a prime. Those corresponding to the lower components of the wave functions, either from initial or final states, are denoted by an underline. Thus the Dirac wave function given in Eq.(\ref{dirac-spinors}) is symbolically written as 
\begin{equation}
\psi_{n \kappa m}(\vec{x}) = \left( 
\begin{array}{r}
i\ \psi_{n(l1/2)jm_{j}} \\
-\ \psi_{n(\underline{l}1/2)jm_{j}}\\
\end{array}
\right)
\end{equation} 

Combining this with the matrix given in Eq.(\ref{multipole-matrix}) one obtains an expression for the matrix elements of the multipole operators between initial and final nuclear states
\begin{eqnarray}
\langle f|O_{JM}|i \rangle & \equiv & \langle j'm_{j}'| O_{JM} |j m_{j} \rangle \nonumber \\
 & = &\langle n'(l'1/2)j'm'_{j}|d_{1}^{JM}|n(l1/2)jm_{j} \rangle \nonumber \\
 & & + \langle n'(\underline{l}'1/2)j'm'_{j}|d_{2}^{JM}|n(\underline{l}1/2)jm_{j} \rangle \nonumber \\
 & &+i \langle n'(l'1/2)j'm'_{j}|b^{JM}|n(\underline{l}1/2)jm_{j} \rangle \nonumber \\
 & &-i \langle n'(\underline{l}'1/2)j'm'_{j}|b^{JM}|n(l1/2)jm_{j} \rangle \nonumber \\
\end{eqnarray} 
where $O_{JM}$ represents any of the multipole operators occurring in Eq.(\ref{me-int-h}). By applying the Wigner-Eckard theorem to the above matrix elements a reduced formed is obtained.
\begin{eqnarray}
\label{me-multipole}
\langle j'm_{j}'| O_{JM} |j m_{j} \rangle & = & \frac{(-1)^{j-m_{j}}}{\sqrt{2J+1}} \langle j'm'_{j}j-m_{j}|j'jJM\rangle \nonumber \\ 
 & & \times \ \langle j'\|O_{J}\|j \rangle \nonumber \\
\end{eqnarray} 
where the reduced matrix elements of $O_{JM}$ are given by:
\begin{eqnarray}
\label{eq:reduced-me}
\langle j'\|O_{J}\|j \rangle & = & \{ \langle n'(l'1/2)j'\|d_{1}^{J}\|n(l1/2)j \rangle \nonumber \\
 & & + \langle n'(\underline{l}'1/2)j'\|d_{2}^{J}\|n(\underline{l}1/2)j \rangle \nonumber \\
 & & + i ( \langle n'(l'1/2)j'\|b^{J}\|n(\underline{l}1/2)j \rangle \nonumber \\
 & & - i \langle n'(\underline{l}'1/2)j'\|b^{J}\|n(l1/2)j \rangle ) \} \nonumber \\
\end{eqnarray} 
Each of these reduced matrix elements is composed of a coefficient corresponding to the angular momentum structure of the matrix element and a radial integral involving the radial wave functions, $G(r)$ and $F(r)$ and spherical Bessel functions. The angular momentum coefficients are tabulated and can be found in~\cite{donnelly-1,donnelly-2}. The radial integrals were performed numerically.

Table~\ref{tb:matrix} shows the form of each matrix element of every multipole operator defined by Eqs.(\ref{eq:multipole-M})-(\ref{eq:multipole-T-mag}). Here we have defined $\mu(k)=k(\lambda_{p}-\lambda_{n})/2m$ and $\eta(k)=1-k^{2}/(m_{\pi}^{2}+q^{2})$, where $q^{2}=k^{2}-\omega_{0}^{2}$ and $k=|\vec{q}|$. The quantity $\omega_{0}$ is identified as the total decay energy.
\begin{table*}
\caption{\label{tb:matrix} This table shows the matrix elements for each of the multipole operators. Here we have used the notation $\mu(k)=k(\lambda_{p}-\lambda_{n})/2m$ and $\eta(k)=1-k^{2}/(m_{\pi}^{2}+q^{2})$, where $q^{2}=k^{2}-\omega_{0}^{2}$ and $k=|\vec{q}|$. Here $\omega_{0}$ is the total decay energy.}
\begin{ruledtabular}
\begin{tabular}{c|c|c|c}
Multipole Operator & $d_{1}^{JM}$ & $d_{2}^{JM}$ & $b^{JM}$ \\
\hline
$M_{JM}$ & $M_{J}^{M}$ & $d_{1}^{JM}$ & $-i\mu(k)\Sigma_{\ J}^{''M}$ \\
$M^{5}_{JM}$ & $-i(\omega_{0}/k)[1-\eta(k)]F_{A}\,\Sigma_{\ J}^{''M}$ & $d_{1}^{JM}$ & $F_{A}\,M_{J}^{M}$ \\
$L_{JM}$ & $0$ & $d_{1}^{JM}$ & $i\Sigma_{\ J}^{''M}$ \\
$L^{5}_{JM}$ & $i\eta(k)F_{A}\,\Sigma_{\ J}^{''M}$ & $d_{1}^{JM}$ & $0$ \\
$T^{\mathrm{el}}_{JM}$ & $\mu(k)\Sigma_{J}^{M}$ & $-d_{1}^{JM}$ & $i\Sigma_{J}^{'M}$ \\
$T^{\mathrm{el},5}_{JM}$ & $iF_{A}\,\Sigma_{J}^{'M}$ & $d_{1}^{JM}$ & $0$ \\
$T^{\mathrm{mag}}_{JM}$ & $i\mu(k)\Sigma_{J}^{'M}$ & $-d_{1}^{JM}$ & $\Sigma_{J}^{M}$ \\
$T^{\mathrm{mag},5}_{JM}$ & $\eta(k)F_{A}\,\Sigma_{J}^{M}$ & $d_{1}^{JM}$ & $0$ \\
\end{tabular}
\end{ruledtabular}
\end{table*}

Several new definitions taken from~\cite{donnelly-1} have been used in this table. They are reproduced here for completeness.
\begin{equation}
\label{eq:newdef1}
\begin{array}{l}
\Sigma_{J}^{M}(k\vec{x}) = \vec{M}_{JJ}^{M}(k\vec{x}) \cdot \vec{\sigma} \\
\Sigma'^{M}_{J}(k\vec{x}) = -i \left \{ \frac{1}{k} \vec{\nabla} \times \vec{M}_{JJ}^{M}(k\vec{x}) \right \} \\
\Sigma ''^{M}_{J} = \left \{ \frac{1}{k} \vec{\nabla} M_{j}^{M}(k\vec{x}) \right \} \cdot \vec{\sigma}\\
M_{J}^{M}(k\vec{x}) \\
\end{array}
\end{equation}
and where 
\begin{equation}
\label{eq:newdef2}
\begin{array}{ccc}
M_{J}^{M}(k\vec{x}) & = & j_{J}(kx)Y_{JM}(\Omega_{x}) \\
\vec{M}_{JL}^{M}(k\vec{x})& = & j_{J}(kx)\vec{Y}^{M}_{JL1}(\Omega_{x}) \\ 
\end{array}
\end{equation}

All reduced matrix elements of the multipole operators are now given by combining Eq.(\ref{eq:reduced-me}) and the expressions given in Table~\ref{tb:matrix}. From the derivation given in~\cite{waleckabook}, one arrives at the following expression which is \textit{general for any semi-leptonic process}.
\begin{eqnarray}
\label{me-general}
\lefteqn{\frac{1}{2j+1}\sum_{m'_{j}} \sum_{m_{j}} |\langle f|H_{\mathrm{w}}|i \rangle|^{2} = \frac{G^{2}}{2} \frac{4\pi}{2j+1} } \nonumber \\
& & \times \{ \sum_{J\geq 1} [ ( |\langle j'\|\mathcal{T}_{J}^{\mathrm{el}}\| j \rangle|^{2} \nonumber \\
& & + |\langle j'\|\mathcal{T}_{J}^{\mathrm{mag}}\| j \rangle|^{2} ) \frac{1}{2} ( \vec{l} \cdot \vec{l}^{*}- l_{3} \cdot l_{3}^{*}) \nonumber \\
& & -\frac{1}{2} i ( \vec{l} \times \vec{l}^{*} ) _{3}2Re ( \langle j'\|\mathcal{T}_{J}^{\mathrm{el}}\|j \rangle \langle j'\| \mathcal{T}_{J}^{\mathrm{mag}}\| j \rangle^{*} )] \nonumber \\
& & + \sum_{J\geq 0} [ l_{3}l_{3}^{*}|\langle j'\| \mathcal{L}_{J}\| j \rangle|^{2} +  l_{0}l_{0}^{*}|\langle j'\| \mathcal{M}_{J}\| j \rangle|^{2} \nonumber \\
& & -2Re ( l_{3}l_{0}^{*} \langle j'\| \mathcal{L}_{J}\| j \rangle \langle j'\| \mathcal{M}_{J}\| j \rangle ^{*} ) ] \} \nonumber \\
\end{eqnarray}
The rest of this derivation, and the calculations done, concentrate on $\beta$-transition processes. For this particular semi-leptonic process the transition rate is given by
\begin{eqnarray}
d\omega & = & \frac{V^{2}}{(2\pi)^{5}}d\Omega_{\epsilon} d\Omega_{\nu} k\epsilon (\omega_{0}-\epsilon)^{2}d\epsilon \frac{1}{2j+1} \nonumber \\ 
 & & \mathop{\sum_{\rm{lepton}}}_{\rm{spins}} \sum_{m'_{j}} \sum_{m_{j}} |\langle f|H_{\mathrm{w}}|i\rangle|^{2} \nonumber \\
\end{eqnarray}
where the $\epsilon$ and $\nu$ subscripts identify quantities related to the electron and the neutrino, respectively, and $\omega_{0}$ is the total decay energy. In addition, V corresponds to the volume of quantization for the lepton wave functions, $\Omega$ is the solid angle in the direction of emission of the lepton, either electron or neutrino, $2j+1$ is the statistical factor corresponding to the initial nuclear state and $m_{j}$ corresponds to the projections of the total angular momentum of the initial and final nuclear states. Using Eq.(\ref{me-general}) and evaluating the spin sums (i.e. lepton traces) one arrives at the final expression that describes $\beta$-transition rates~\cite{waleckabook}.
\begin{eqnarray}
\label{main-eq}
\lefteqn{d\omega = \frac{V^{2}}{(2\pi)^{5}}d\Omega_{\epsilon} d\Omega_{\nu} k\epsilon (\omega_{0}-\epsilon)^{2}d\epsilon \frac{4\pi G^{2}}{2j+1}} \nonumber \\
 & & \{ \sum_{J\geq 1} [ ( |\langle j' \| \mathcal{T}_{J}^{\mathrm{el}} \| j \rangle |^{2} + |\langle j' \| \mathcal{T}_{J}^{\mathrm{mag}} \| j \rangle |^{2})( 1-(\hat{q} \cdot \vec{\beta})(\hat{q} \cdot \hat{\nu})) \nonumber \\
 & & \mbox{} + \hat{q} \cdot (\hat{\nu} - \vec{\beta}) 2Re( \langle j' \| \mathcal{T}_{J}^{\mathrm{el}} \| j \rangle \langle j' \| \mathcal{T}_{J}^{\mathrm{mag}} \| j \rangle ^{*} )] \nonumber \\
 & & + \sum_{J\geq 0} [ (1-\hat{\nu} \cdot \vec{\beta}+2(\hat{q} \cdot \vec{\beta})(\hat{q} \cdot \hat{\nu})) | \langle j' \| \mathcal{L}_{J} \| j \rangle |^{2} \nonumber \\
 & & + (1+\vec{\beta} \cdot \hat{\nu}) |\langle j' \| \mathcal{M}_{J} \| j \rangle |^{2} \nonumber \\
 & & - \hat{q} \cdot (\hat{\nu}+\vec{\beta}) 2Re(\langle j' \| \mathcal{L}_{J} \| j \rangle \langle j' \| \mathcal{M}_{J} \| j \rangle ^{*} )]\} \nonumber \\
\end{eqnarray}
where $\hat{q}=\vec{q}/|\vec{q}|$, $\hat{\nu}=\vec{\nu}/\nu$ and $\vec{\beta}=\vec{k}/\epsilon$, $\vec{q}$ is the momentum transfer, $\vec{\nu}$ is the momentum of the neutrino and $\nu$ its energy, $\vec{k}$ is the momentum of the electron and $\epsilon$ its energy.

This expression can be integrated over the corresponding phase-space to obtain the desired $\beta$-transition rate. All the nuclear structure input to this formula is embedded in the reduced matrix elements of the multipole operators. These in turn are generally composed of four terms as given by Eq.(\ref{eq:reduced-me}) and the expressions of Table~\ref{tb:matrix}. For the last part of this section a summary of the expressions needed to evaluate these reduced matrix elements is included.

As mentioned above, all reduced matrix elements of the multipole operators are composed of two factors: one corresponding to the angular momentum structure of the matrix elements and the other corresponding to an integral over the initial and final wave functions weighted by spherical Bessel functions. Their expressions are given in~\cite{donnelly-2}.
\begin{eqnarray}
\label{eq:coeff-M}
\lefteqn{\langle n'(l'1/2)j'\|M_{J}(k\vec{x})\|n(l1/2)j \rangle =} \nonumber \\
 & & (4\pi)^{-1/2}A_{J}(l'j';lj)\langle n'l'j'|j_{J}(\rho) |nlj \rangle \nonumber \\
\end{eqnarray}
\begin{eqnarray}
\label{eq:coeff-sigma}
\lefteqn{\langle n'(l'1/2)j'\| \Sigma_{J}(k\vec{x})\|n(l1/2)j \rangle =} \nonumber \\
 & & (4\pi)^{-1/2}D_{J}(l'j';lj)\langle n'l'j'|j_{J}(\rho) |nlj \rangle \nonumber \\
\end{eqnarray}
\begin{eqnarray}
\label{eq:coeff-sigma-p}
\lefteqn{\langle n'(l'1/2)j'\| \Sigma_{J}^{'}(k\vec{x})\|n(l1/2)j \rangle =} \nonumber \\
 & & (4\pi)^{-1/2} \{ -J^{1/2} D_{J}^{+}(l'j';lj)\langle n'l'j'|j_{J+1}(\rho) |nlj \rangle \nonumber \\
 & & +(J+1)^{1/2} D_{J}^{-}(l'j';lj)\langle n'l'j'|j_{J-1}(\rho) |nlj \rangle \} \nonumber \\
\end{eqnarray}

\begin{eqnarray}
\label{eq:coeff-sigma-pp}
\lefteqn{\langle n'(l'1/2)j'\| \Sigma_{J}^{''}(k\vec{x})\|n(l1/2)j \rangle =} \nonumber \\
 & & (4\pi)^{-1/2} \{ (J+1)^{1/2} D_{J}^{+}(l'j';lj)\langle n'l'j'|j_{J+1}(\rho) |nlj \rangle \nonumber \\
 & & +J^{1/2} D_{J}^{-}(l'j';lj)\langle n'l'j'|j_{J-1}(\rho) |nlj \rangle \} \nonumber \\
\end{eqnarray}

In all these expressions $j_{J}(\rho)$ corresponds to a spherical Bessel function of order J and argument $\rho = kx$.
The coefficients $A_{J}$, $D_{J}$, $D_{J}^{+}$ and $D_{J}^{-}$ are tabulated in~\cite{donnelly-1,donnelly-2} for a large, but limited number of transitions. For coefficients of transitions not included in these tables, explicit expressions of the above coefficients in terms of 3-j and 6-j symbols can be found in ~\cite{donnelly-1}~\footnote{Some coefficients for the hole-hole transition $(1\rm{g}_{9/2})^{-1}\rightarrow(1\rm{h}_{11/2})^{-1}$ where obtained using the explicit expressions given in~\cite{donnelly-1} and tabulated 3-j and 6-j coefficients~\cite{3j6jsymbols}.}.
The integral over the initial and final radial wave functions is given by
\begin{equation}
\langle n'l'j'|\theta (\rho) |nlj \rangle = \int_{0}^{\infty} dr \ u(r)_{n'\kappa '} \theta(\rho) v(r)_{n \kappa}
\end{equation}

Here $\theta(\rho)$ stands for the appropriate spherical Bessel function and $u(r)$ and $v(r)$ for either the upper or lower component of the Dirac wave function, i.e. either $G(r)$ or $F(r)$ of Eq.(\ref{dirac-spinors}), of the initial and final nucleon state.

\section{Results}

The calculations of $\beta$-transition rates are done by numerical integration of Eq.(\ref{main-eq}) over the electron and neutrino phase-space. These transition rates are very sensitive to the exact decay energy $\omega_{0}$. Although the results obtained for total binding energies of the nuclei involved in this study agree within 1\% of its experimental values, here the calculations of the decay rates are done using the experimental decay energies so as to take into account the full phase-space available to the process. This way, no additional uncertainties are introduced and the direct contributions coming from the calculated nuclear wave functions can be accounted for. The decay energies used here are taken from~\cite{expdata-sn133,expdata-in131}. The Dirac wave functions used in these calculations are solutions to the Kohn-Sham equations derived from the effective lagrangian given in~\cite{furnstahl} using the G1 parameter set.

\begin{figure}
\vspace{1cm}
\includegraphics[width=0.45\textwidth,height=0.45\textwidth]{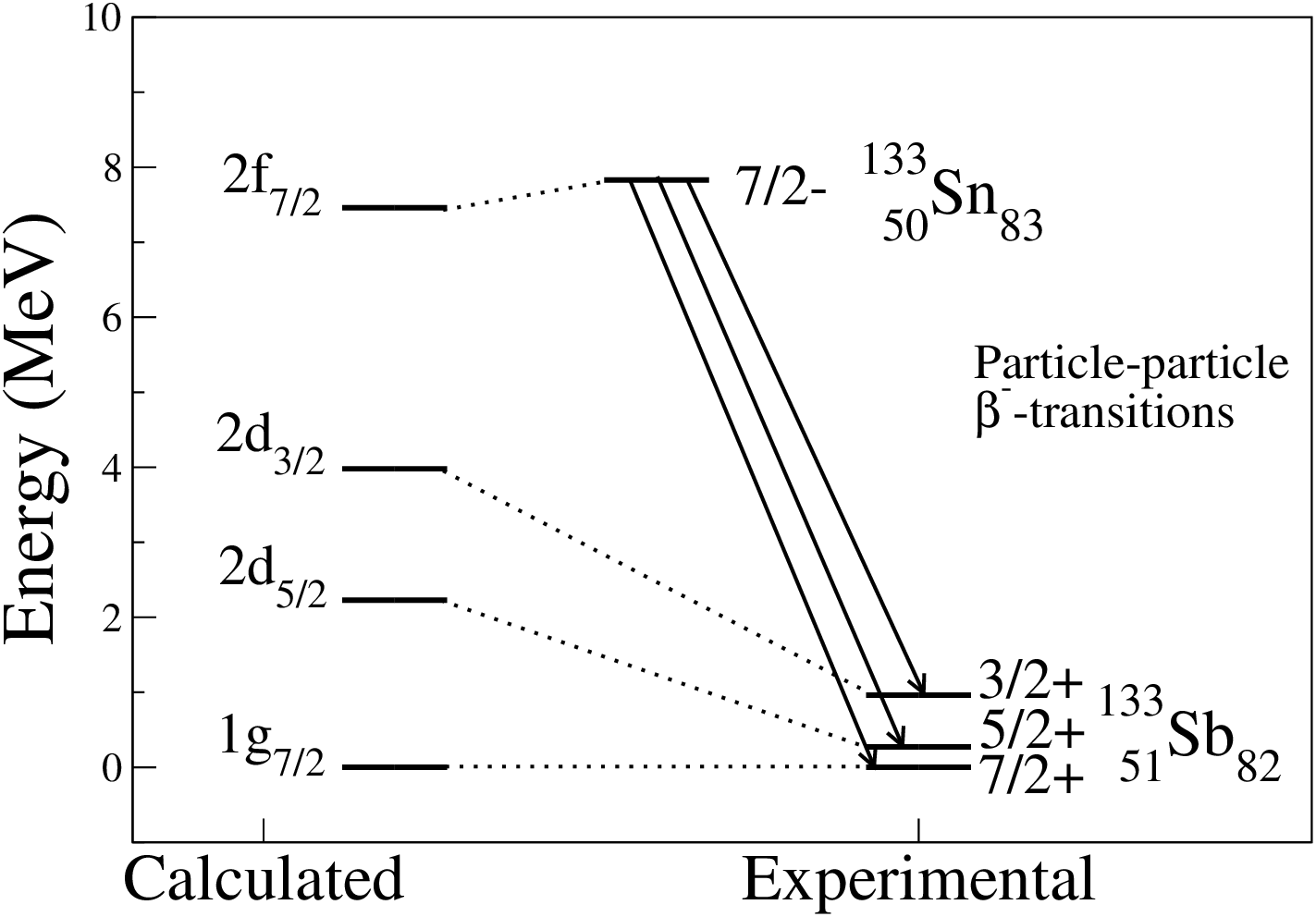}
\caption{\label{fig:pp-levels} Single-particle level spectrum of parent and daughter nuclei in the case of particle-particle transitions. All energy levels are measured from the calculated or experimental ground-state level, accordingly.
\vspace{1mm}}
\end{figure}

\begin{figure}
\includegraphics[width=0.45\textwidth,height=0.45\textwidth]{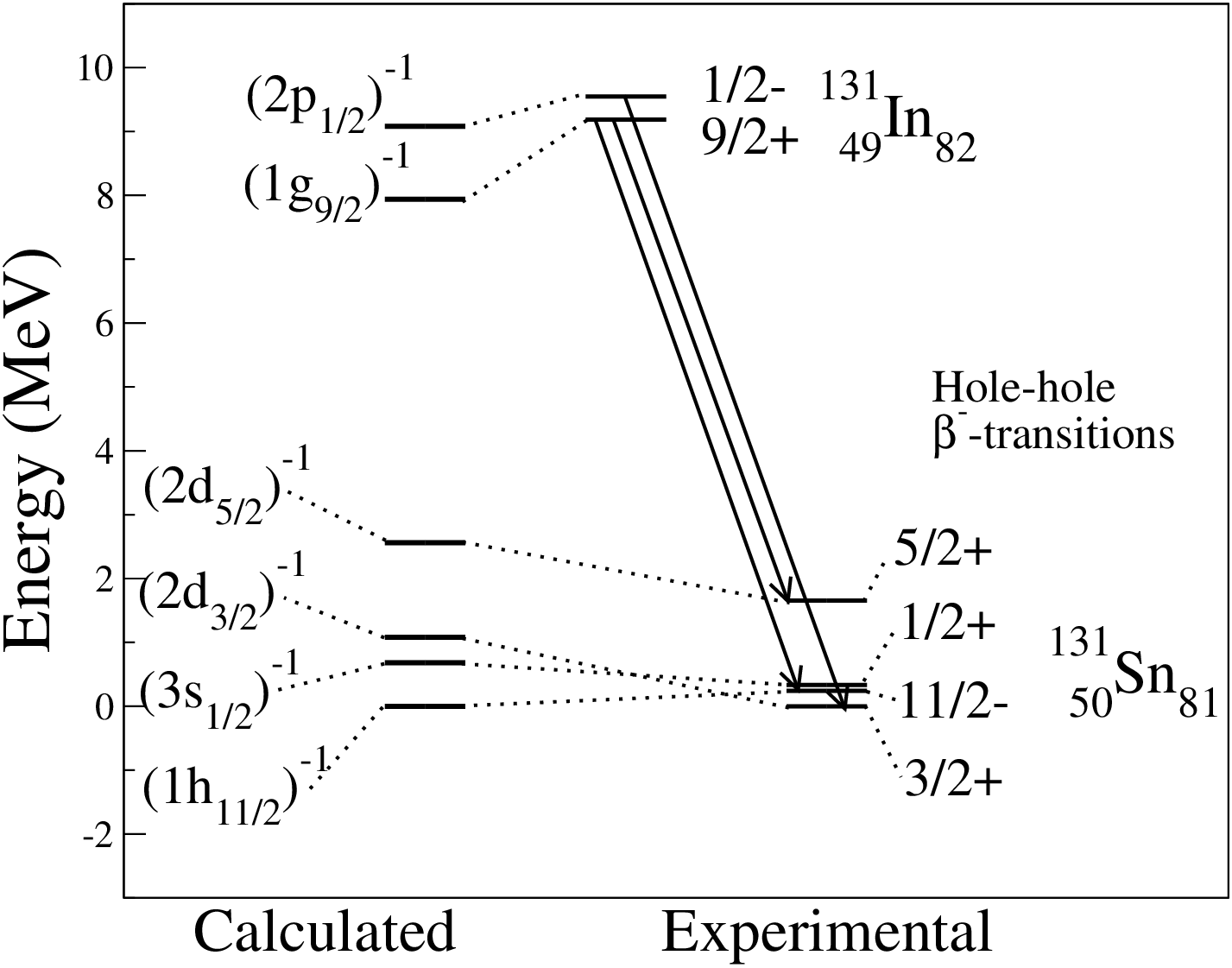}
\caption{\label{fig:hh-levels} Single-hole level spectrum of parent and daughter nuclei in the case of hole-hole transitions. All energy levels are measured from the calculated or experimental ground-state level, accordingly.
\vspace{0.9cm}}
\end{figure}

Two types of transitions are investigated in this paper: 1) particle-particle and 2) hole-hole. The diagrammatic representation of the transitions studied here are shown in Fig.~\ref{fig:pp-levels} for transitions of the particle-particle type and in Fig.~\ref{fig:hh-levels} for the hole-hole type. The level structure indicated on the left side of each figure corresponds to the results obtained using the EFT/DFT approach. On the right are the measured levels. In both cases the energy levels are measured with respect to the ground-state level, either the calculated (left) or experimental (right). The results obtained give, for the nuclei considered in the particle-particle transitions, the correct  level ordering although not the right splitting. This is in accord with DFT with regard to excited states. On the other hand, based on the agreement obtained for the ground-state binding energies of the parent and daughter nuclei~\cite{huertas}, it is expected that the wave functions corresponding to these states, i.e. 2f$_{7/2}$ and 1g$_{7/2}$ respectively, are a good approximation. These particles are outside a filled core and in this approach they only interact with it through the mean-fields. An assumption has been made here that these nuclei, with one particle outside its core, can be described by spherically symmetric wave functions. 

In the case of the nuclei considered here for hole-hole transitions, several complications emerge. First, the level ordering of the daughter nucleus, $^{131}_{\ 50}\textrm{Sn}_{81}$ is reproduced except for the (2d$_{3/2}$)$^{-1}$ state, which is the measured ground-state. The calculated ground-state corresponds to the 1h$_{11/2}$ level, which lays approximately 200 keV off the experimental value. Another complication arises from the fact that there are no ground-state to ground-state transitions in the hole-hole case, so a direct comparison with the particle-particle situation cannot be made. Here all transitions go from or to excited states. Yet another complication comes from the assumption that it is possible to approximate a hole state wave function by a single-particle wave function. The hole state corresponds to an unfilled core, with one particle missing, and no interactions among the particles in the unfilled core have been taken into account beyond the mean-field one. These three situations make it difficult to compare the hole-hole type of transitions with experiment. Still the results, as will be shown below, are not so different than other calculations and in general follow the pattern of behavior of the experimental transition rates.

The calculation of $\beta$-transition rates using Eq.(\ref{main-eq}) employs the initial and final nuclear wave functions. These wave functions enter into the calculation in a different form depending on the type of transition, i.e. particle-particle or hole-hole. In the particle-particle case the initial and final wave functions correspond with the initial and final nuclear states. For example, for the ground-state to ground-state transition $^{133}_{\ 50}\textrm{Sn}_{83} \rightarrow ^{133}_{\ 51}\textrm{Sb}_{82}$ the initial and final wave function are 2f$_{7/2}$ and 1g$_{7/2}$ respectively, see Fig.~\ref{fig:pp-levels}. Hole-hole transitions, on the contrary, can be interpreted as a particle transitions going in the opposite direction~\cite{electron-scattering}. Thus, for example, in the transition $^{131}_{\ 49}\textrm{In}_{82} \rightarrow ^{131}_{\ 50}\textrm{Sn}_{81}$ going from ground-state to the (1h$_{11/2}$)$^{-1}$ state, the initial and final wave functions would be (1h$_{11/2}$) and (1g$_{9/2}$) respectively~\footnote{In the hole-hole case, there is an additional phase factor coming from exchanging the creation and destruction operators in the matrix elements of the tensor operator. The states are characterized by their total angular momentum quantum number $j$, therefor, if the initial and final states are denoted by $| j_{b}^{-1} \rangle$ and $| j_{a}^{-1} \rangle$, then the matrix elements of the transition matrix element of a tensor operator $O_{JM}$ will be given by : $\langle j_{a}^{-1}\|O_{J}\|j_{b}^{-1} \rangle = -(-1)^{j_{\rm{a}}+j_{\rm{b}}-J} \langle j_{b}\|O_{J}\|j_{a} \rangle $~\cite{electron-scattering}.}.

\begin{figure}
\vspace{1cm}
\includegraphics[width=0.45\textwidth,height=0.45\textwidth]{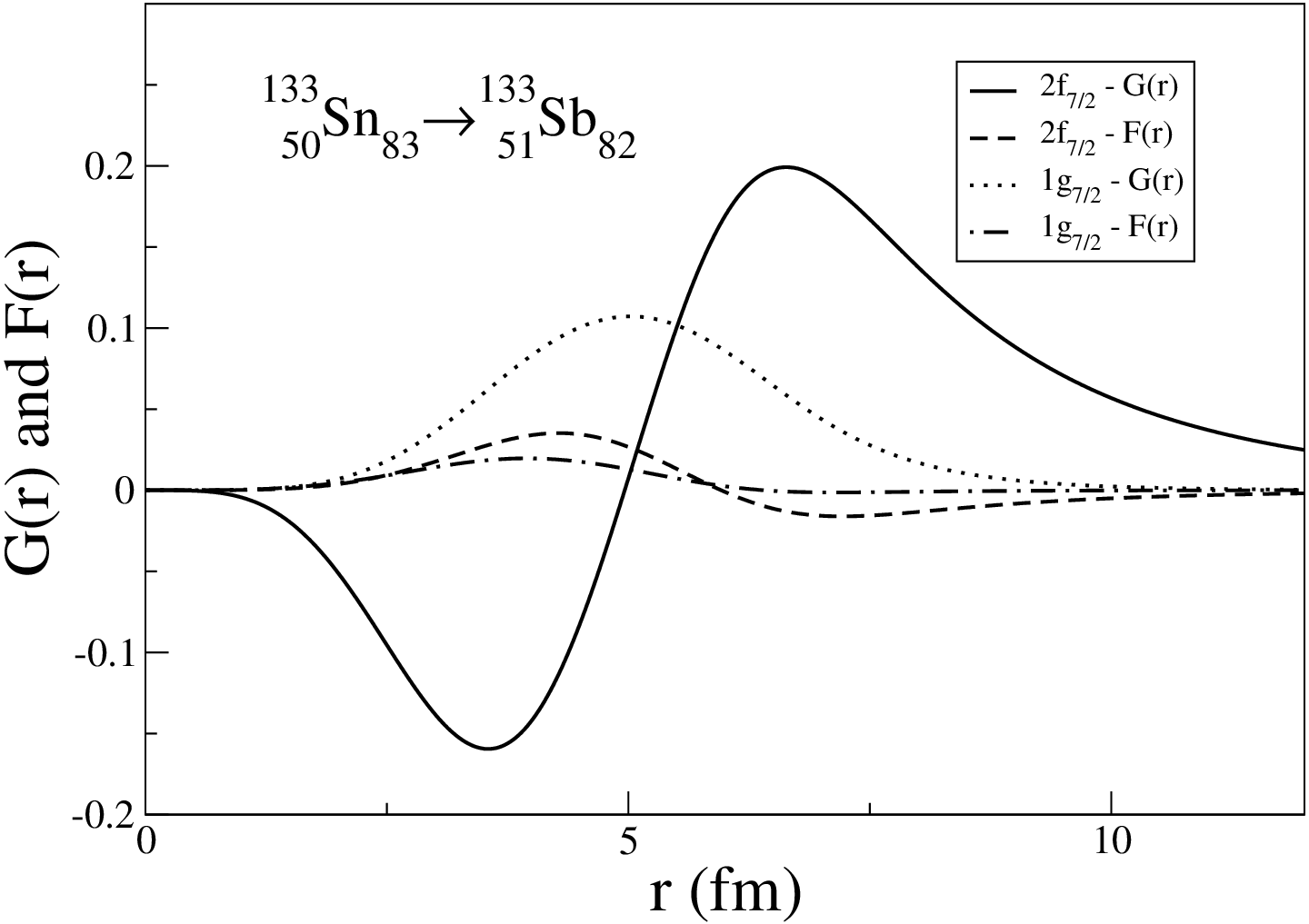}
\caption{\label{fig:wf-pp-gstogs}Upper and lower components of the initial and final nucleon wave functions. This corresponds to a particle-particle transition proceeding from the ground-state of $^{133}_{\ 50}\textrm{Sn}_{83}$ to the ground-state of $^{133}_{\ 51}\textrm{Sb}_{82}$.
\vspace{1cm}}
\end{figure}

\begin{figure}
\includegraphics[width=0.45\textwidth,height=0.45\textwidth]{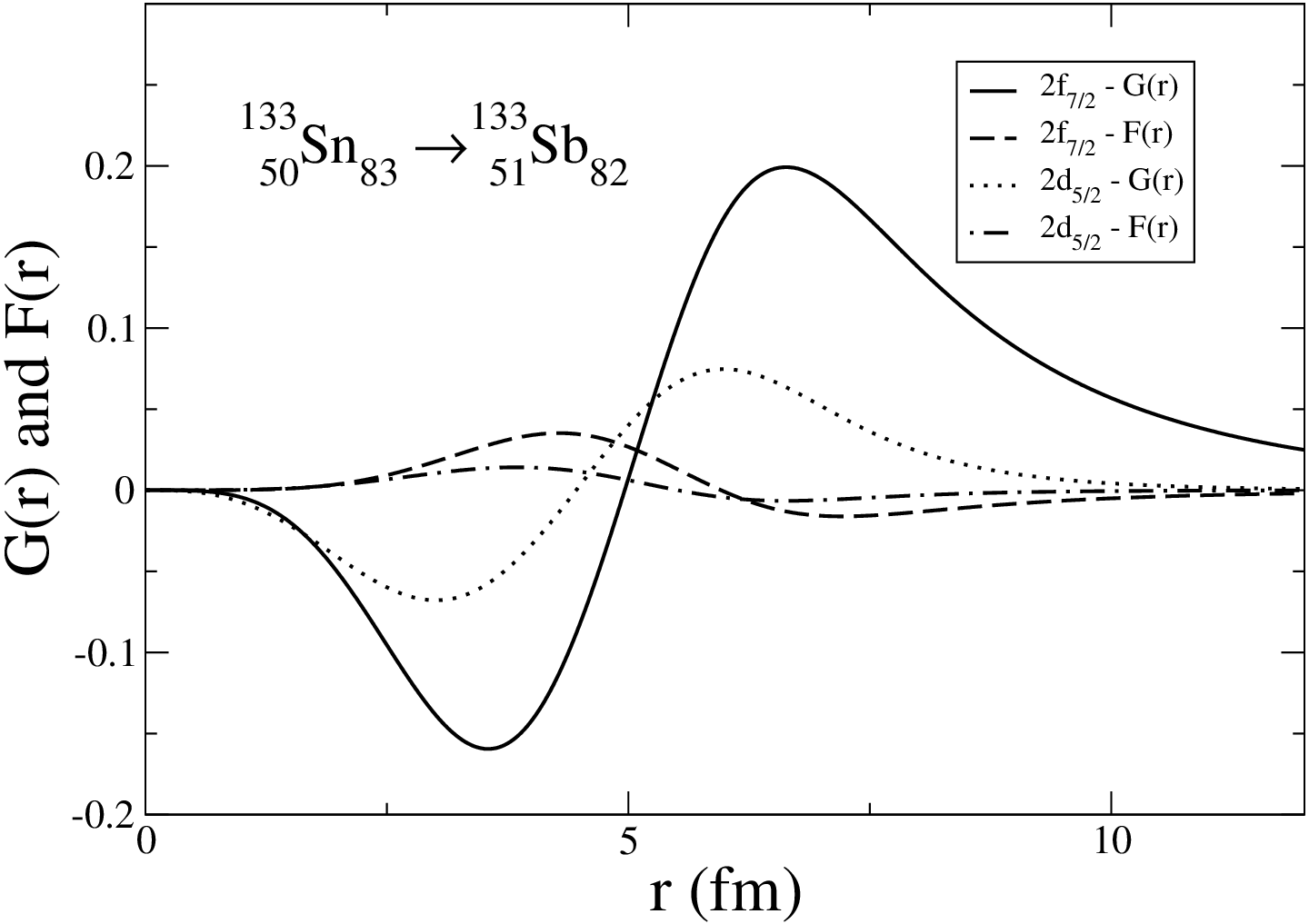}
\caption{\label{fig:wf-pp-gsto2d5_2}Upper and lower components of the initial and final nucleon wave functions. This corresponds to a particle-particle transition proceeding from the ground-state of $^{133}_{\ 50}\textrm{Sn}_{83}$ to an excited state of $^{133}_{\ 51}\textrm{Sb}_{82}$.}
\end{figure}

\begin{figure}
\vspace{1cm}
\includegraphics[width=0.45\textwidth,height=0.45\textwidth]{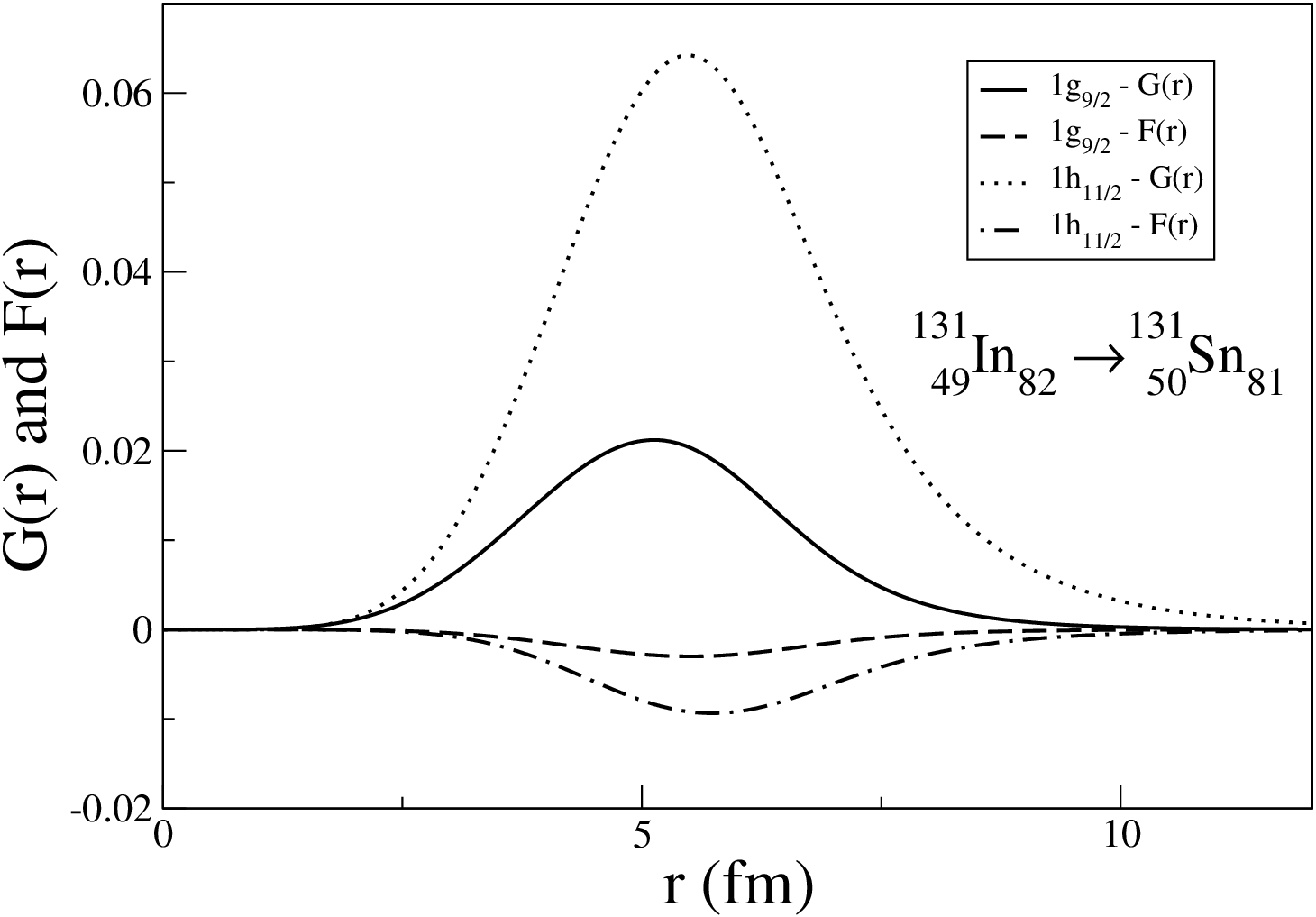}
\caption{\label{fig:wf-hh-gsto1h11_2}Upper and lower components of the initial and final nucleon wave functions. This corresponds to a hole-hole transition proceeding from the ground-state of $^{131}_{\ 49}\textrm{In}_{82}$ to the \textit{calculated} ground-state of $^{131}_{\ 50}\textrm{Sn}_{81}$.\vspace{1cm}}\end{figure}

\begin{figure}
\includegraphics[width=0.45\textwidth,height=0.45\textwidth]{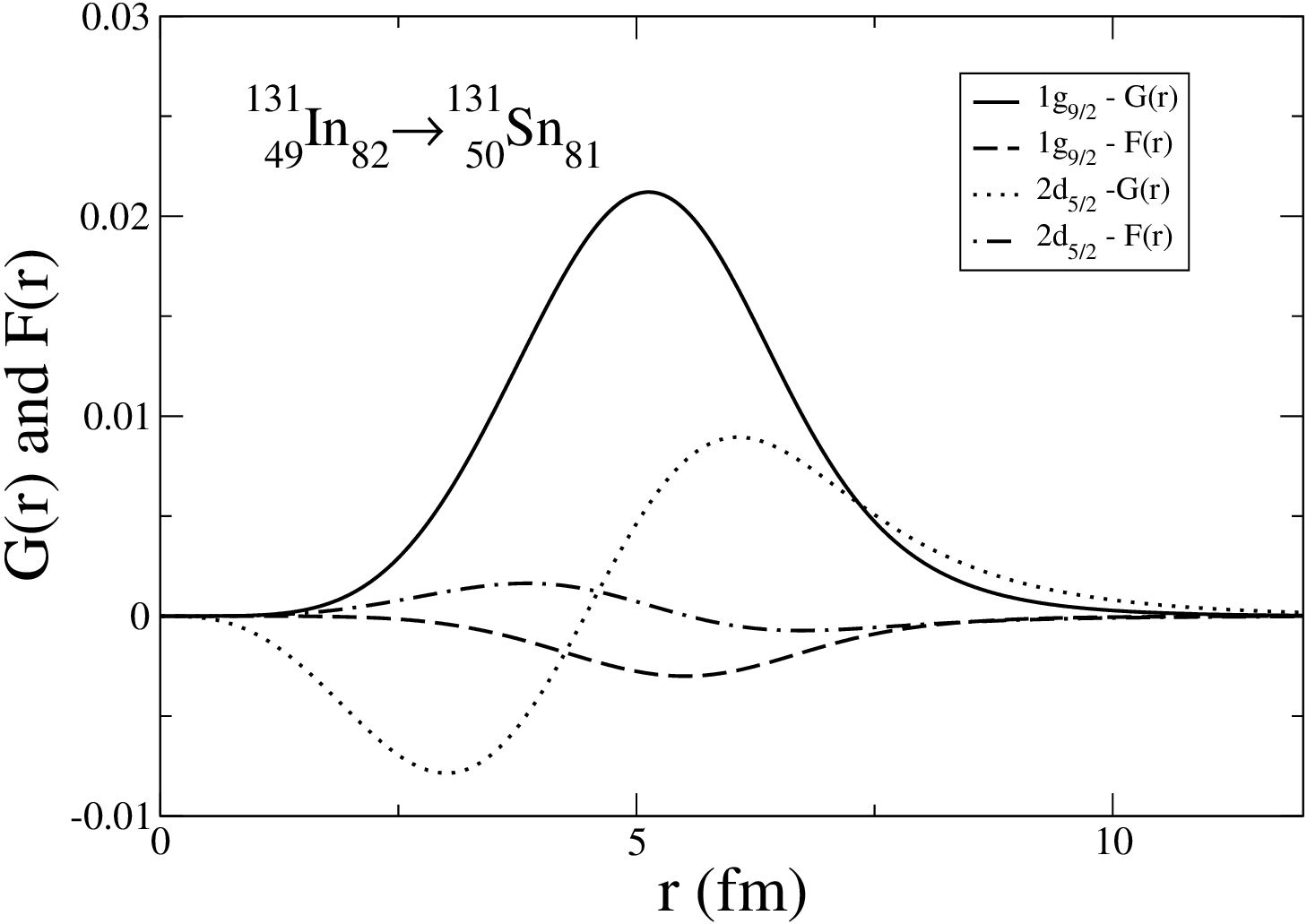}
\caption{\label{fig:wf-hh-gsto2d5_2}Upper and lower components of the initial and final nucleon wave functions. This corresponds to a hole-hole transition proceeding from the ground-state of $^{131}_{\ 49}\textrm{In}_{82}$ to an excited state of $^{131}_{\ 50}\textrm{Sn}_{81}$.}
\end{figure}

Examples of the single-particle wave functions used in the calculation of $\beta$-transition rates are given in Fig.~\ref{fig:wf-pp-gstogs} through~\ref{fig:wf-hh-gsto2d5_2}. In these figures the upper and lower components of the Dirac wave functions, i.e. the $G(r)$ and $F(r)$ functions, are plotted simultaneously as a function of the radial distance r, in Fermis.

\begin{figure}
\vspace{1.9 cm}
\includegraphics[width=0.45\textwidth,height=0.45\textwidth]{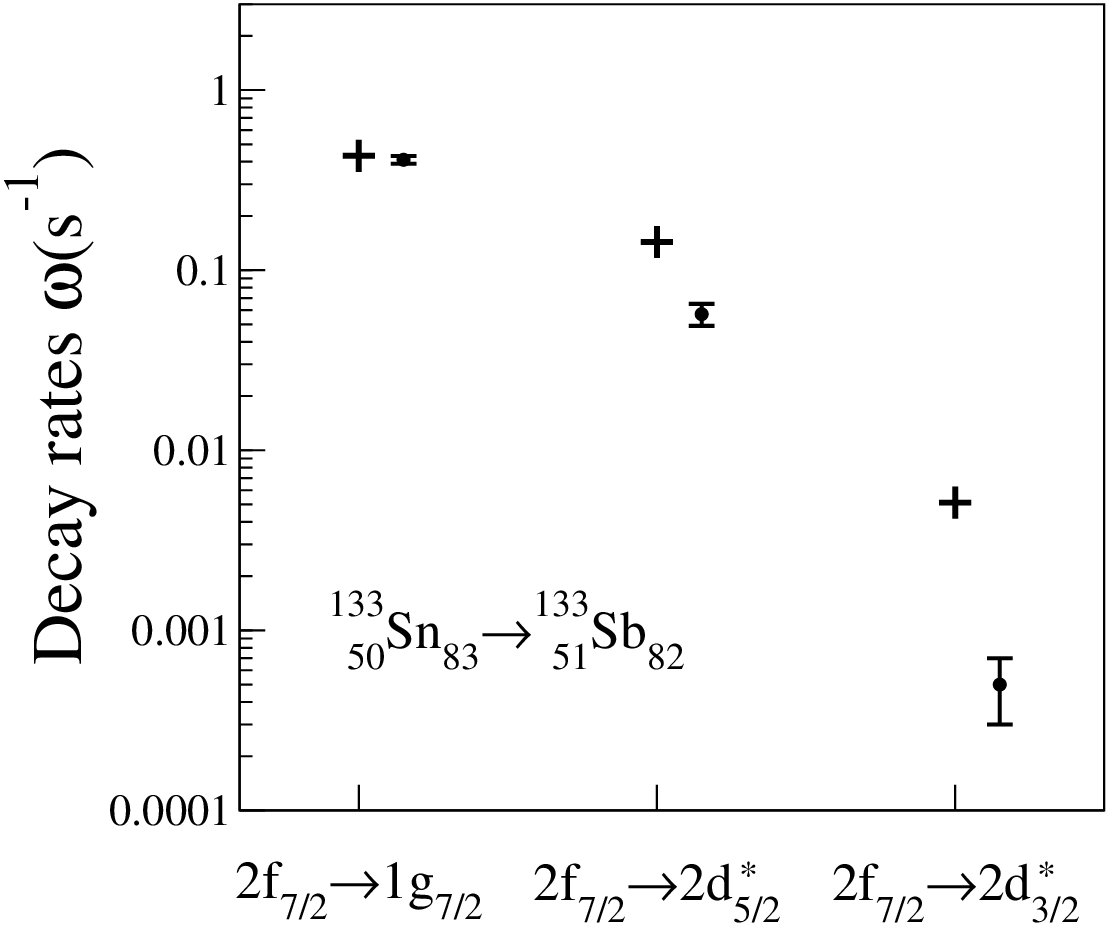}
\caption{\label{fig:pp-results} This figure shows the calculated results of the $\beta$-transition rates in the cases of particle-particle transitions and the experimental values.}
\end{figure}

\begin{figure}
\vspace{-0. cm}
\includegraphics[width=0.45\textwidth,height=0.45\textwidth]{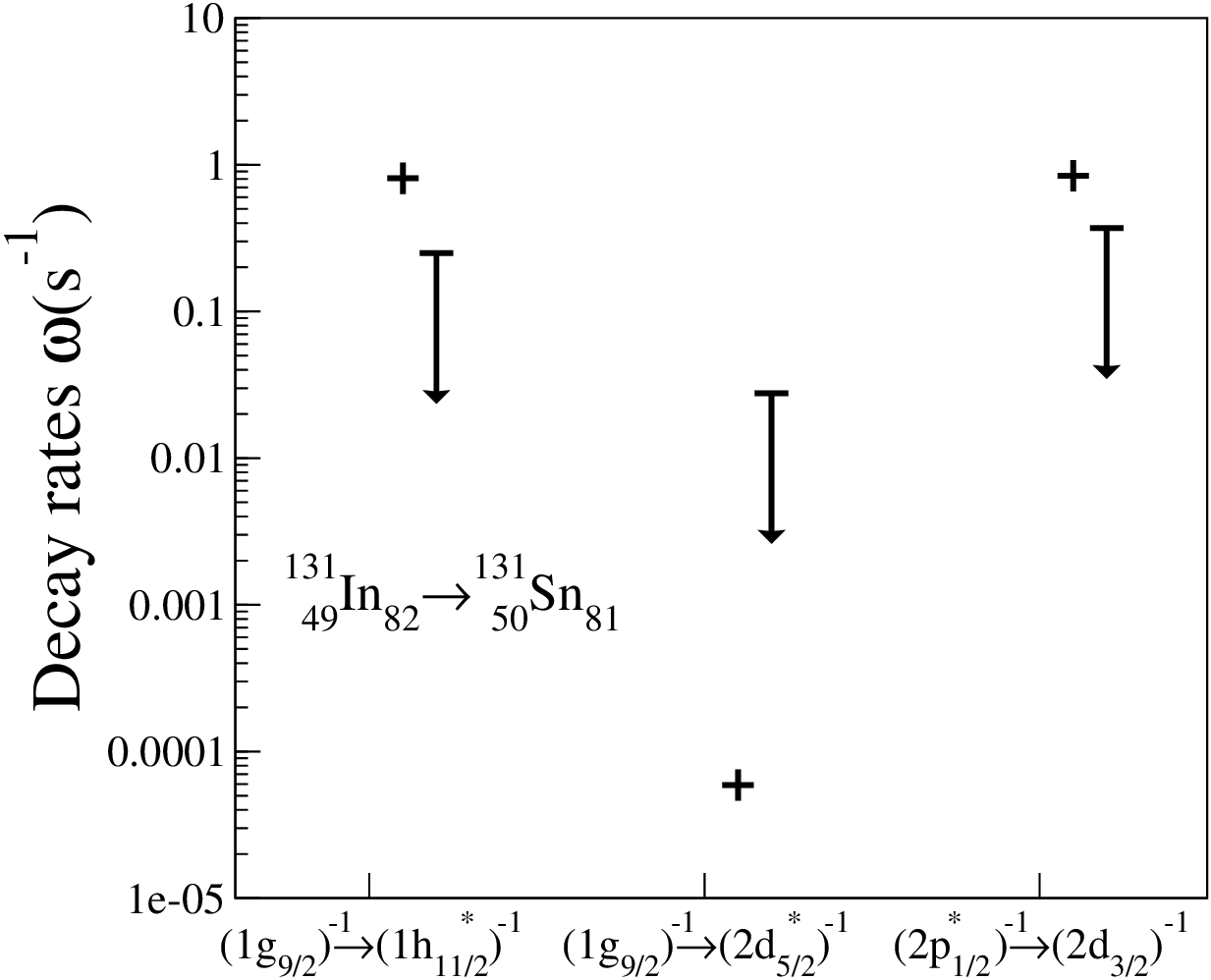}
\caption{\label{fig:hh-results}This figure shows the calculated results of the $\beta$-transition rates in the cases of hole-hole transitions and the experimental values.}
\end{figure}

The results of the calculation of $\beta$-transition 
rates for the particle-particle type are shown in Fig.~\ref{fig:pp-results}. These are compared with experimental values for each of the transitions. Here the calculated results are indicated with a cross. It can be seen that \textit{the ground-state to ground-state transition is well reproduced}. The ratio of calculated to experimental values is 1.05, or a deviation of 5\%. The other transitions go from the ground-state of the parent nucleus to an excited state in the daughter nucleus. 

The results for these transitions show systematically a larger value of the decay rates compared with the experimental ones. This agrees well with what is expected of DFT. Since the calculations have been performed using the experimental decay energies, the only input from the EFT/DFT approach at this point has been the single-particle Dirac wave functions. These, on the other hand, are used to construct the nuclear ground-state densities and since this quantity is in principle well reproduced, the result of the ground-state to ground-state transition serves as an indirect way of checking it.

\begin{figure}
\vspace{-0.7 cm}
\includegraphics[width=0.45\textwidth,height=0.45\textwidth]{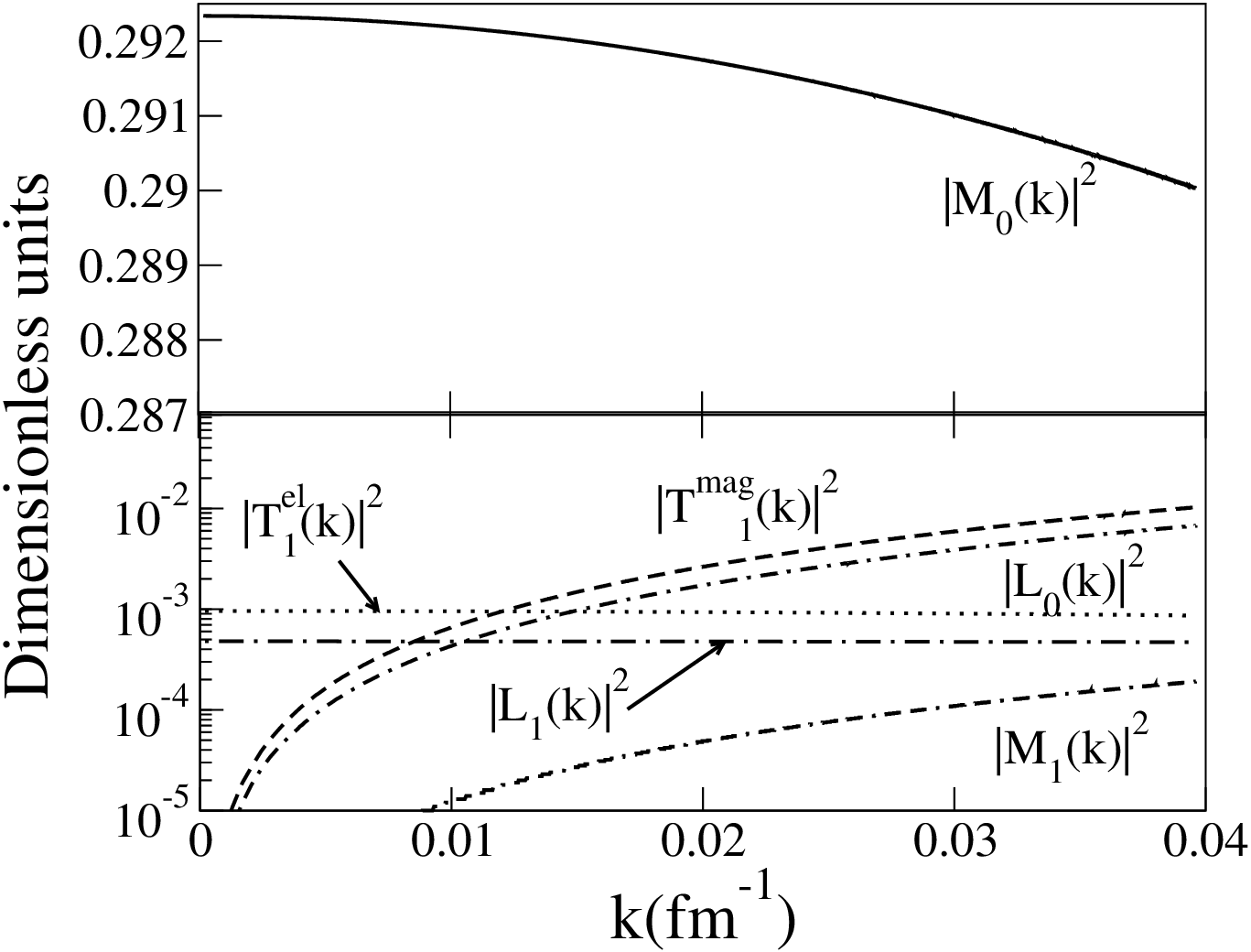}
\caption{\label{fig:multipoles}Magnitude of the multipoles defined in Eqs. (\ref{eq:multipole-M})-(\ref{eq:multipole-T-mag}) for J=0 and J=1. These multipoles are the dominant terms contributing to the particle-particle transition undergoing between the ground-states of $^{133}_{\ 50}\rm{Sn}_{83}$ and $^{133}_{\ 51}\rm{Sb}_{82}$. In this figure $\rm{M}_{J} \equiv \mathcal{M}_{J}$, $\rm{L}_{J} \equiv \mathcal{L}_{J}$, $\rm{T}^{\rm{el}}_{J} \equiv \mathcal{T}^{\rm{el}}_{J}$ and $\rm{T}^{\rm{mag}}_{J} \equiv \mathcal{T}^{\rm{mag}}_{J}$.}
\end{figure}

Returning to the results for the ground-state to ground-state transition in the particle-particle case, Fig.~\ref{fig:multipoles} shows the dominant multipoles as a function of the $k=|\vec{q}|$. The values ploted correspond to the square of the magnitud of the multipoles defined in Eqs. (\ref{eq:multipole-M})-(\ref{eq:multipole-T-mag}). The multipoles ploted correspond to values of $J=0$ and $J=1$, with $|\mathcal{M}_{0}(k)|^{2}$ being the most dominant term. From this figure it is also clear that at low momentum transfer the only non-vanishing multipoles are $|\mathcal{M}_{0}(k)|^{2}$, $|\mathcal{T}^{\rm{el}}_{1}(k)|^{2}$ and  $|\mathcal{L}_{1}(k)|^{2}$, in accordance with the analysis made in~\cite{waleckabook}.

The results obtained in the hole-hole type of transitions are shown in Fig.~\ref{fig:hh-results}. In this case, the results are less satisfactory than in the particle-particle case. The experimental results are shown here as arrows indicating upper limits to the decay rates. From the three cases calculated here, only one lies within the experimentally determined range. This transition corresponds to a ground-state to excited state hole-hole transition. The two other cases shown occur from ground-state to excited state and excited to ground-state respectively. Their numerical results are very similar. In these cases, there is no direct ground-state to ground-state transition which makes it difficult to compare with the particle-particle case. 

\section{Conclusions}

This paper shows the results of calculations of the $\beta$-transition rates of nuclei close to $^{132}_{\ 50}$Sn$_{182}$. These calculations were inspired from the success of the results obtained in a previous paper~\cite{huertas} dealing with the application of the EFT/DFT approach to the nuclear many-body system to nuclei far from stability. In reaching this goal several important results have been obtained.

\begin{enumerate}
\item A general expression to calculate any semi-leptonic process has been derived. The main equations and definitions needed are given by Eqs. (\ref{eq:reduced-me}), (\ref{eq:newdef1})-(\ref{eq:newdef2}), (\ref{me-general}), (\ref{eq:coeff-M})-(\ref{eq:coeff-sigma-pp}) and Table~\ref{tb:matrix}. The nucleon wave functions are the Dirac wave functions obtained from solving the Kohn-Sham equations derived in the EFT/DFT approach and considers relativistic corrections to all orders in the nucleon wave functions. Additionally, the electroweak currents used correspond to the leading Noether currents obtained directly from the same effective lagrangian. In this sense, the calculation is self-consistent since all relevant elements have been obtained directly from a single theory.

\item The results of the $\beta$-transition rates calculated using this formalism agree within 5\% with the experimental values for the transition going from ground-state to the ground-state in the particle-particle case; see Fig.~\ref{fig:pp-results}. DFT can reproduce ground-state observables and since the results of the total binding energy of the parent and daughter nuclei essentially agree with experiment, it would be expected that the last particle in the system is well described. The fact that the $\beta$-transition rates deviate systematically when the transitions go to excited states gives additional support to this view.

\item The results of the $\beta$-transition rates in the case of hole-hole transitions are less accurate than the particle-particle case; see Fig.~\ref{fig:hh-results}. Here the absence of a pure ground-state to ground-state transition does not allow a better assessment of the argument given above for the particle-particle case. The fact that the true ground-state of $^{131}\textrm{Sn}$ is not reproduced makes it even more difficult, although the results obtained show at least the same pattern of behavior as the experimental data. 
\end{enumerate}

I like to thank Dr. J. D. Walecka for his encouragement, support and advice. This work is supported in part by DOE grant DE-FG02-97ER41023.


\newpage

\end{document}